\documentclass[10pt,twocolumn,amsmath,amssymb,aps,superscriptaddress,prb]{revtex4}

\usepackage{graphicx}       
\usepackage{dcolumn}        
\usepackage{bm}             
\usepackage{amsmath}        
\usepackage{color}          

\definecolor{parchment}{rgb}{.9,.9,.6}

\newcommand{\bq}{\bm q}
\newcommand{\bQ}{\bm Q}
\newcommand{\bk}{\bm k}
\newcommand{\br}{\bm r}
\newcommand{\bR}{\bm R}
\newcommand{\bs}{\bm s}
\newcommand{\btau} {\bm \tau}
\newcommand{\bdelta} {\bm \delta}
\begin{document}
\title{Theory of Graphene Chiral Quasiparticle LDOS maps}
\author{T. Pereg-Barnea and A.H. MacDonald}
\affiliation{Department of Physics, University of Texas at Austin, Austin TX 78712-1081, USA}
\date{\today}

\begin{abstract}
We present a theory of momentum space local density-of-states (LDOS) maps $N(\bq,\omega)$
in graphene.  The LDOS map has both intravalley contributions centered near zero momentum and
reciprocal lattice vectors and intervalley contributions displaced by the
wavevector ${\bm K'-\bm K}$ which connects graphene's two distinct Dirac points.
Using graphene's Dirac equation chiral quasiparticle continuum model, we obtain
analytic results which explain the qualitative differences between these two LDOS map features.
We comment on the sensitivity of both $N(\bq,\omega)$ features to the mix of atomic length
scale and smooth disorder sources present in a particular graphene sample.
\end{abstract}

\maketitle

\section{Introduction}
Graphene, a honeycomb lattice of carbon atoms, is a recently realized\cite{GeimDiscovery} two dimensional (2D) electron system (2DES)
with a variety of unique properties.\cite{GeimReview,GeimPT}  This intriguing electronic system is
now being actively explored both experimentally and
theoretically.  Graphene is described at low energies by a 2D massless-Dirac wave equation in which the role of
spin is assummed by a pseudospin which represents the two atoms in its unit cell.  Many of the unusual properties
of graphene sheets, including the shift in the densities at which quantum Hall\cite{GeimQH,ColumbiaQH} pleateaus occur,
follow from momentum-space Berry phases\cite{KaneSHE,SinitsynSHE} associated the pseudospin (sublattice) degree-of-freedom.
Graphene quasiparticles have definite pseudospin
chirality, {\em i.e.} definite projection of pseudospin along momentum measured from one
of the two independent Brillouin-zone
corner points at which the gap vanishes.  The experimental observation of the quantum
Hall effect in graphene was an important demonstration that these 2DESs behave, at least in some respects,
very nearly in the ideal way anticipated by Wallace\cite{wallace} many years ago in his early analysis of the electronic structure of
graphite.  Wallace realized that the states near the Fermi level of a graphene sheets should
separate carbon $\pi$-orbital bonding and antibonding bands, and that the gap between these bands would vanish
at two-points in the honeycomb lattice Brillouin-zone, the Dirac points of graphene.
Experiments have confirmed this simple picture in most respects,
something that was not {\em a priori} obvious given the potential for either electron-electron
interactions or disorder to alter physical properties.  Indeed the quantum Hall effect, which can be viewed as a
topological property of the 2DES, tends to be forgiving on details.  In view of the tremendous interest in studying graphene sheets and
characterizing their disorder, there is strong motivation for expanding the comparison between experiment
and theoretical analysis based on Wallace's $\pi$-band model to new observables.

In this article we present theoretical predictions for the local density-of-states of weakly
disordered graphene sheets, a quantity which can
be measured using scanning-tunneling-microscopy (STM).
A new appreciation of the ability of this type of measurement to shed light on
the character of the disorder in a sample, and also on underlying clean system
electronic properties, has emerged from a highly successful series of studies
of cuprate superconductors\cite{HTCexperiments}.  The experiments rely on the ability to make stable
atomic-resolution STM scans of the LDOS as a function of energy $\omega$ over a large real-space field of view.
(The energy $\omega$ is varied by
changing the bias voltage between the STM tip and the sample.)
Preliminary experimental data on graphene sheets is already available\cite{experiments}, and we can expect that further refinements in sample quality and experimental technique will enable detailed analysis which will extract much useful information. Previous theoretical work\cite{BenaKivelson,Bena,Wang,Cheianov1,Cheianov2} has discussed the numerical construction of
LDOS maps from graphene's honeycomb lattice $\pi$-band tight-binding model.
Analytic expressions for the LDOS modulations were obtained by Bena\cite{Bena} through an expansion in powers of $1/r$ of the amplitudes in real space.  Peres {\it et al.} have looked at impurity induced localized states at low energies in the $T$-matrix approximation\cite{Peres}.
In this theoretical contribution we focus on the Dirac equation
continuum limit of the $\pi$-band model, from which it is possible to obtain analytic results which we believe
can contribute to a more meaningful interpretation of experimental LDOS maps.

In this paper we suggest that it can be useful to measure A and B sublattice LDOS maps separately, something which is possible in principle since the experiments have atomic resolution.  As we explain, the difference signal (which will normally be weak) and the sum signal are complementary probes of a sample and its disorder.  With this method
one is able to extract more information and consequently identify the type of disorder in the system.

Because of the disorder always present in real materials the LDOS never has the lattice periodicity.
When Fourier transformed to momentum space, the additional modulations\cite{HTCtheory} reveal the
dominant momentum transfers between quasiparticle states at energy $\omega$ (measured from the Dirac point) and hence map out the electronic structure. In a superconductor the Bogoliubov-deGennes coherence factors, which specify the particle and hole amplitudes in
BCS quasiparticles, help determine the relative probability of scattering between
different constant energy surface segments\cite{PBF1,PBF2}.
As we shall see shortly, the pseudospin mixing present in graphene quasiparticle excitations
plays a similar role.  At small $\omega$ the momentum-transfers in graphene are naturally classified as
intra-valley (small momentum transfer close to the same Brillouin-zone corner) or inter-valley (large
momentum transfer from one valley to the other.)   We will see that both intra-valley and
inter-valley scattering amplitudes are influenced by pseudospin chirality, but
because the two valleys have opposite chirality their STM momentum-space maps differ
qualitatively.

Our paper is organized as follows.  In Section II we briefly summarize the weak-scattering analysis which
has been used to provide STM momentum-space maps with a simple and consequential interpretation.  In Sections III and IV we
apply the analysis first to a graphene $\pi$-orbital tight-binding model and then to the
low energy Dirac equation continuum model.  The Dirac equation model allows many elements of the
calculation to be carried out analytically, enabling us to provide more guidance on the
qualitative interpretation of STM momentum space maps.  We conclude in Section V with a brief summary and
discussion.

\section{The Weak Disorder Approximation}

\subsection{Real-space LDOS maps}

The influence of a scattering potential $V(\br)$ on the LDOS of an otherwise clean system can be
described by expanding the Greens' function in powers of $V(\br)$.
LDOS maps are most revealing when the scattering potential, $V(\br)$, is weak, justifying truncation
at first order - the Born approximation.
For a simple parabolic band, this approximate treatment leads to the well known Freidel
oscillations\cite{Friedel}.  In a LDOS-map experiment, weak disorder provides an
electronic system with its own weakly-coupled probe - one which is able to provide
momentum-resolution of single-particle properties.  It is generally assumed that the weak-coupling
approximation is at least qualitatively valid  whenever sharp
features appear experimentally in the momentum-space maps we describe below.  Maps similar
to the ones we calculate later in this paper should emerge from
LDOS-map studies of graphene if samples of sufficiently high quality can be prepared.

To model graphene we consider the following perturbed Hamiltonian:
\begin{eqnarray}
{\cal H} = {\cal H}_0 + {\cal H}_v \nonumber \\
{\cal H}_v = \sum_{\br} V(\br)n(\br)
\end{eqnarray}
where the unperturbed Hamiltonian ${\cal H}_0$ is either
the $\pi$-band tight binding\cite{wallace,TB} model for graphene or its Dirac continuum limit.\cite{linearized}
The Dyson equation for the Greens' function is then
\begin{eqnarray}\label{eq:Dyson}
G(\br,\br',\omega) && = \; \; G^0(\br'-\br,\omega) \nonumber \\
&+& \int d\bs \; G^0(\bs-\br,\omega)V(\bs)G(\bs,\br',\omega),
\end{eqnarray}
where $G^0(\br'-\br,\omega)$ is the unperturbed Greens' function of a clean-limit graphene quasiparticle at energy $\hbar\omega$ and
$G(\br,\br',\omega)$ is the inhomogeneous disordered system Greens' function
which includes the effect of scattering by the potential $V(\br)$.
In order to describe a non-Bravais lattice, it is convenient to write the Hamiltonian and therefore the Greens function as $2\times2$ matrices with the argument $\br$ defined only on the Bravais lattice. We define the vector $\br$ to lie on the A sublattice and the B sublattice atoms are obtain through a shift by the vector $\btau$.
For on-site disorder $V$ is diagonal and its elements specify the A and B site potentials in a particular unit cell.  It is convenient to write
\begin{equation}
V(\br) \equiv V_0(\br)\, \sigma_0 + V_3(\br) \, \sigma_3,
\end{equation}
(Here $V_0 + V_3 = V_A$, $V_0 - V_3 = V_B$, $\sigma_0$ is the $2 \times 2$ identity matrix and $\sigma_3$ is the diagonal Pauli matrix.)  Disorder sources which are smooth on an atomic scale, for
example smooth ripples\cite{Katsnelson,deJuan,ripples} or Coulomb potentials from
remote ionized impurities\cite{Nomura,DasSarma,CastroNetoCoulombImp}, will
contribute only to $V_0$, whereas atomic scale disorder sources\cite{Castronetodisorder}
like impurity atoms or vacancies, will have large differences between $V_A$ and $V_B$ and
contribute to both $V_0$ and $V_3$.
The local density of states at position $\br$ and energy $\hbar\omega$, $N(\br,\omega)$, is given by
\begin{eqnarray}\label{eq:LDOS}
N(\br,\omega) = -{1 \over \pi}\times \begin{cases} {\rm Im}[ G_{AA}(\br,\br,\omega)] &  {\rm on\; A\; site} \\
                                          {\rm Im}[ G_{BB}(\br,\br,\omega)] & {\rm on\; B\; site}
                                                \end{cases}.
\end{eqnarray}

\subsection{Momentum-space LDOS maps}

The zeroth order term in the potential expansion of the Greens' function gives the clean-limit LDOS
which is periodic and therefore has non-zero Fourier components only at reciprocal lattice vectors.
The disorder-induced non-periodic spatial modulations appear in the subsequent terms.
The Born approximation sums over all quantum paths which include a single scattering event.
In terms of the exact single-particle eigenstates of a disordered graphene system, the Born approximation to the LDOS
accounts for the corrections to the Bloch wavefunctions of the perfect crystal which
appear at first order when disorder is treated as a perturbation.
In principle one can continue the perturbation series to include higher orders of the potential.
For a single-scatterer, the full series is easily summed to construct the full $T$-matrix\cite{Bena,balatsky}, a necessity
when strong scattering leads to impurity resonance states.  In this paper,
motivated in part by the absence of any experimental evidence for resonances in typical graphene sheets and
also by the possibility of extracting more information about the sample from STM experiments when this limit applies,
we concentrate on the weak disorder limit.

A STM measurement naturally projects a quantum wavefunction to one sublattice or the other
and not, as implicitly assumed by Bena and Kivelson\cite{BenaKivelson}, to a $\bm k$-dependent Bloch pseudospinor.  This point is discussed in detail by Bena and Montambaux\cite{BenaMontambaux}.
We restrict our attention, without loss of generality, to evaluating the LDOS measured on the A sublattice.
This sublattice separation will allow distinction between the two types of scattering profiles as will be demonstrated shortly.
In the Born approximation, we may restrict our attention to a single-site impurity potential which we take
to be located in the unit cell with lattice vector ${\bf L}=0$; the distributed potential
case can be constructed simply by adding contributions from different unit cells.
The first order correction in the Greens' function is\cite{footnote}
\begin{eqnarray}
G^{(1)}_{AA}(\br,\br,\omega) = [G^0(\br-0,\omega)V(0)\sigma_{0/3}G^0(0-\br,\omega)]_{AA} = \nonumber \\
V(0)[G^0_{AA}(\br)G^0_{AA}(-\br) \pm G^0_{AB}(\br)G^0_{BA}(-\br)].
\label{ba}
\end{eqnarray}
In Eq.(\ref{ba}) the first term describes the case of measuring and scattering on the same sublattice,
and the second term the case of measuring on one sublattice and scattering from the other.

Electronic structure information is revealed most directly by Fourier transforms of the LDOS map.
We evaluate the momentum-space LDOS map first from graphene's $\pi$-orbital tight-binding model and later take the
continuum limit of these calculations to obtain analytic Dirac-model results.
The tight-binding model Bloch Hamiltonian
\begin{eqnarray}
\label{tbm}
{\cal H}_{0\bk} = \begin{pmatrix} 0 & \gamma_{\bk} \\ \gamma^\dagger_{\bk} & 0 \end{pmatrix} \nonumber \\
\gamma_{\bk} = t(e^{i k_y / \sqrt{3}} + 2e^{i k_y / 2\sqrt{3}}\cos(k_x/ 2)),
\end{eqnarray}
where $t$ is the hopping amplitude and we measure length in units of the triangular lattice constant (not the carbon-carbon distance)
a=$2.46\AA$.  The $2\times 2$ unperturbed Matsubara Greens' function is then $(i\omega-{\cal H})^{-1}$.

The tight-binding model Hamiltonian (Eq.(\ref{tbm})), and hence
$G^0$, is constructed in a representation in which
Bloch state basis functions have a phase difference $\bk\cdot\btau$ between site 2 and site 1 in
every unit cell.  Here $\btau$ is the vector from site A to site B.
It is critical that the same gauge is used when the
Fourier transform of the LDOS is constructed, in particular in evaluating the
Fourier transform of the LDOS
on one sublattice due to scattering from a site on the other:
\begin{eqnarray}\label{eq:ABBA}
\sum_{\bR} e^{-i\bq\bR} G^0_{AB}(\bR,\omega)G^0_{BA}(-\bR,\omega) = \nonumber \\
\sum_{\bR}e^{-i\bR\bq}\!\!\!\sum_{\bk,\bk'\in {\rm BZ}}\!\!\!\!e^{i\bk (\bR + \btau)}G^0_{AB}(\bk,\omega)e^{-i\bk' (\bR + \btau)}G^0_{BA}(\bk',\omega) \nonumber\\
=\sum_{\bk\in {\rm BZ}}e^{i(\bQ-\bq) \cdot \btau}G^0_{AB}(\bk,\omega)G^0_{BA}(\bk-\bq+\bQ,\omega).
\end{eqnarray}
The $\exp(i(\bk-\bk')\cdot \btau)$ in the second line results from the gauge choice in which the origin is taken to be on the A sublattice.
In the last line we have noted that the lattice vector sum yields a $\delta$-function which restricts $\bk-\bk'$ to be equal to $\bq$ up to a reciprocal lattice vector $\bQ$.  $\bQ$ is therefore chosen such that $\bk-\bq+\bQ$ is in the momentum-space primitive cell.  The full Fourier transformed LDOS is therefore given by:
\begin{eqnarray}\label{eq:FTLDOS}
\delta N^A(\bq,\omega) &=& {1 \over 2 \pi i}\!\!\sum_{\mu = 0,3}\!\!V_\mu(\bq)(\Lambda_\mu(\bq,\omega-i\delta) - \Lambda_\mu(\bq,\omega+i\delta)) \nonumber \\
\Lambda_\mu(\bq,i\omega) &=&\sum_{\bk} [G^0_{AA}(\bk,i\omega)G^0_{AA}(\bk-\bq,i\omega)\\ \nonumber
&\pm& e^{i\bQ\btau}G^0_{AB}(\bk,i\omega)G^0_{BA}(\bk-\bq+\bQ,i\omega)]
\end{eqnarray}
where the retarded and advanced parts of $\Lambda$ are obtained by analytically continuing $i\omega \to \omega \pm i\delta$ and $V_\mu(\bq)$ is the Fourier transformed potential.
Note that the factor $e^{-i\bq\btau}$ which appears in Eq.(\ref{eq:ABBA}) is not present in Eq.(\ref{eq:FTLDOS}).  This is due to a 'form factor' which is implicit in the Fourier transform of the potential $V(\bq)$, i.e., due to the distance $\btau$ between the two atoms in the unit cell a phase of $e^{i\bq\btau}$ appears in the B-component of the potential.
These expressions do not include the $\pi$-orbital form factor which is expected to gradually decrease momentum-space amplitudes at wavevectors outside the Brillouin zone.

It is possible to describe any arrangement of impurities from Eq.(\ref{eq:FTLDOS}) by combining the results for $\sigma_0$ and $\sigma_3$ potentials.  For example, a single impurity on the A sublattice is represented by $\sigma_0+\sigma_3$ (and therefore setting $V_0 = V_3$).  It is also possible to obtain the amplitude of the LDOS modulations on the B sublattice from Eq.(\ref{eq:FTLDOS}) by replacing labels $A \leftrightarrow B$ and reversing the direction of the basis vector, $\btau \to -\btau$.  The single impurity results are given by:
\begin{eqnarray}
\Lambda_A^A(\bq) = {1\over 2}(\Lambda_0^A(\bq)+\Lambda_3^A(\bq)) \nonumber \\
\Lambda_B^A(\bq) = {1\over 2}(\Lambda_0^A(\bq)-\Lambda_3^A(\bq)) \nonumber \\
\Lambda_A^B(\bq) = {1\over 2}(\Lambda_0^B+\Lambda_3^B) = {1\over 2}((\Lambda_0^{A})^*-(\Lambda_3^A)^*)\nonumber \\
\Lambda_B^B(\bq) = {1\over 2}(\Lambda_0^B-\Lambda_3^B) = {1\over 2}((\Lambda_0^A)^*+(\Lambda_3^A)^*)
\end{eqnarray}
where the subscript represents the position of the impurity and the superscript is the sublattice on which the measurement is done.

In this work we have assumed that the experimental data is separated to the two sublattices before the Fourier transform is performed.  If the experimental data is Fourier transformed without the sublattice separation the two diagonal components of the perturbed Greens' function should be added $(G_{AA}+G_{BB})$.   In the case of a single impurity (as studied by Bena\cite{Bena}) the result obtained in this way is identical to our $\delta N^A_0(q)$.
In the case of two identical impurities on the A and B sites, the FT constructed from both the A and B sites ($\delta N_0^{A+B}$) is equal to the real part of our $\delta N_0^A$.  In the case of two equal and opposite impurities ($\sigma_3$ scattering) the combination of A and B sublattices cancels the real part and the resulting $\delta N_3^{A+B}(q)$ is purely imaginary and is equal to the imaginary part of our $\delta N_3^A(q)$.  These identifications of real and imaginary parts apply, of course, when only one unit cell of the lattice has scatterers, and the A site of this sublattice is chosen as the origin of coordinates.  In the more general case of distributed disorder, which we expect applies to real materials, it is nevertheless true that the difference between A and B sublattice maps arises purely from the $\Lambda_3$ response whereas the sum of the A and B maps will be dominated by the $\Lambda_0$ response if the dominant disorder varies smoothly on an atomic scale.

\section{Tight-binding Model Numerical Results}
\begin{figure}[h]
\includegraphics[width=\columnwidth]{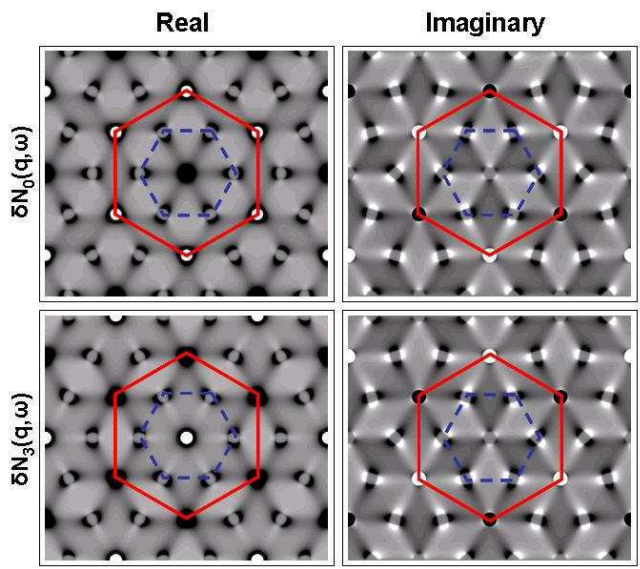}
\caption{The real and imaginary parts of $\delta N_{0/3}(\bq)$ represented in grey-scale (dark is high) in
momentum-space.  The LDOS maps were constructed from graphene's $\pi$-orbital
tight-binding model.   These numerical results were obtained
at energy $\omega= 0.2 t$.  The solid lines (red online) are a guide to
the eye connecting reciprocal lattice vectors.  The dashed lines (blue online)
are a guide to the eye connecting inter-valley features in the
LDOS map.  The dashed lines also form the triangular lattice
Brillouin-zone boundary.}
\label{fig:TB}
\end{figure}
Momentum-space LDOS maps constructed by evaluating Eq.(\ref{eq:FTLDOS})
from the full tight binding Hamiltonian are illustrated in Fig.(\ref{fig:TB}).
These results are for identical amplitude scatterers, the $\delta N_{0}(\bq)$ result, and
opposite scatterers, the $\delta N_{3}(\bq)$ results, in the unit cell at the origin.
Results for a general scatterer can be obtained by inserting its Fourier transform
in Eq.(\ref{eq:FTLDOS}).
Efficient evaluation of the LDOS is achieved by Fourier transforming the unperturbed Greens function to real space (through the fast Fourier transform algorithm), multiplying matrices appropriately,
and then taking the imaginary part.
This quantity is then Fourier transformed back to momentum space, mimicking the procedure
used to analyze experimental data.
A closer zoom on intra- and inter-valley features from Fig.(\ref{fig:TB}) is given in the inset of Fig.(\ref{fig:zero}) and in Fig.({\ref{fig:zoom}) respectively.

These results are filled with interesting features
which reflect the physics of graphene sheets, motivating the experimental studies which
this work aims to assist.  Most obvious is the expected appearance of clear separate features
associated with intra-valley and inter-valley scattering.  Bloch states near the Fermi energy
of neutral graphene sheets appear close to the two valley points, $K = (4\pi/3,0)$ and $K' = (8\pi/3,0)$
at which the $\pi$ bonding and antibonding bands meet.  States with an
energy $\omega$ (measured from the neutral system Fermi level) occur close to a circle
centered on $K$ or $K'$ with radius $k_{\omega}=\omega/v$ where $v$ is graphene's Dirac-cone velocity.
Smooth disorder potentials contribute only to $V_{0}(\bq)$ and have large
amplitudes only for scattering within these valleys.  They therefore
contribute to LDOS map features only near $\bq=0$ or reciprocal lattice vectors.
We see in Fig.(\ref{fig:TB}) that the intra-band $\Lambda_{0}(\bq)$ features at $|\bq| = 2 k_{\omega}$,
associated with scattering across a Dirac cone, are much weaker than the corresponding
features in $\Lambda_{3}(\bq)$.
 We also note that the inter-valley features which
appear in the LDOS map near wavevectors $\pm(K-K)'$ have an interesting angular variation which
is absent in the intra-valley feature.  Neither intra-band or inter-band features are
periodic under translation by a reciprocal lattice vector, as explained previously.
Another feature of the LDOS modulations which appears in the numerical results is the structure in the vicinity of reciprocal lattice vectors, i.e., in the corners of the (red online) solid hexagonal zone in Fig.(\ref{fig:TB}).  The amplitude of the modulations near these points may be obtained from the modulations of small momentum transfer through a $2\pi/3$ rotation of the B-atom scattering contribution.
The above features may be understood in the context of the Dirac model, presented in the next section.
\section{Dirac model analytic results}

\begin{figure}
\includegraphics[width = \columnwidth]{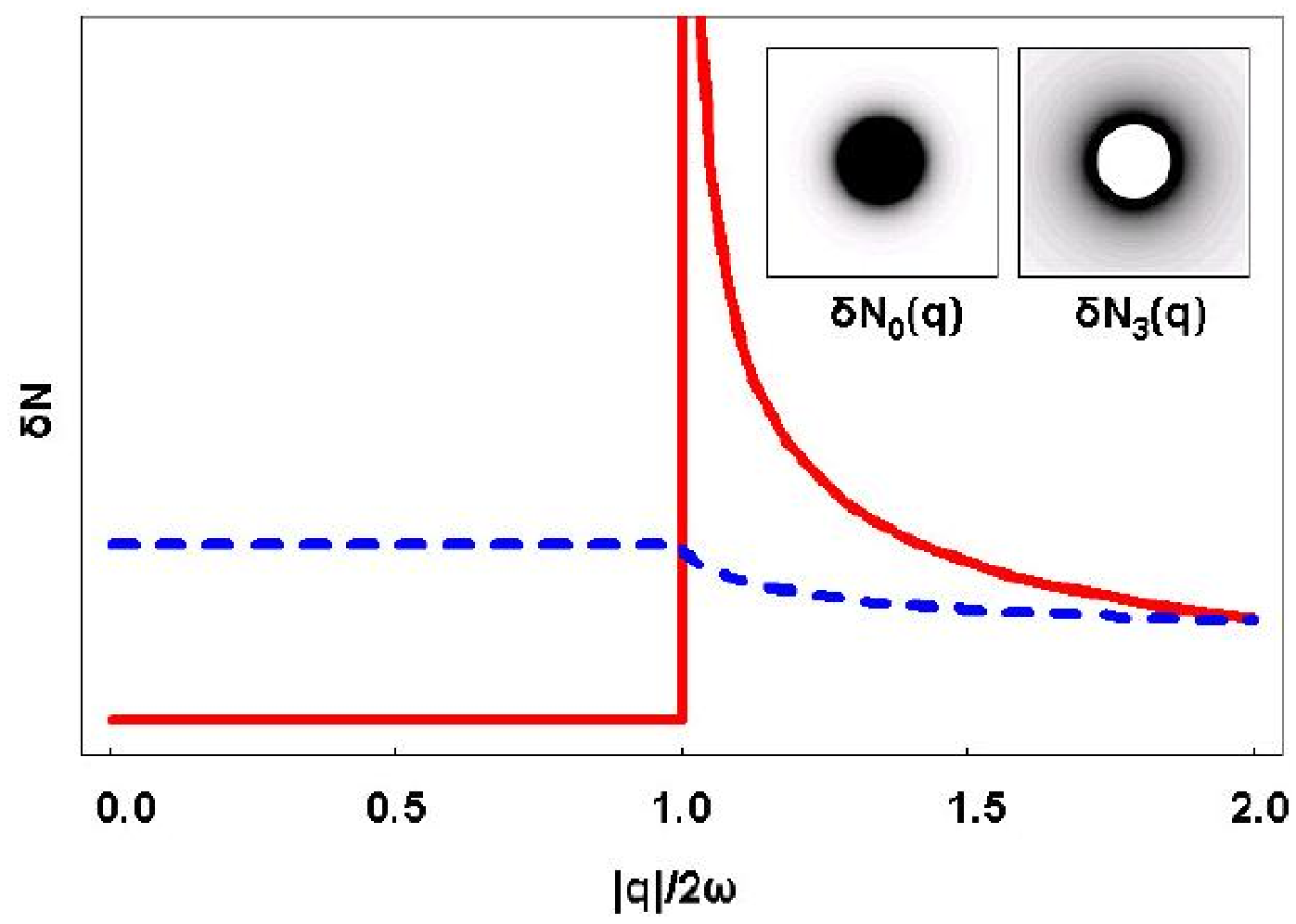}
\caption{FT LDOS of $\Lambda_0$ (dashed, blue online) and $\Lambda_3$ (solid, red online) type scattering, as a function of $|q|$ in unites of $2\omega/v$, around zero momentum transfer.  Note that the FT is purely real
and dependent only on $|\bq|$.  The inset shows the functions in momentum space (to be compared with the numerical data presented in Fig.(\ref{fig:TB}).}\label{fig:zero}
\end{figure}

One of the reasons for the excitement around graphene is its low energy behavior.
At two valley points in the Brillouin zone the energy bands touch and the dispersion is linear.
This makes graphene a zero gap semiconductor whose low energy Hamiltonian is a condensed matter realization of the Dirac model.
Using the massless Dirac model, we can achieve a deeper understanding of the
numerical results reported in Fig.(\ref{fig:TB}).

For our analytic calculations
it is convenient to choose a unit cell in momentum space which
includes both valleys in its interior.  The more symmetric triangular lattice Brillouin zone (BZ)
has the disadvantage of separating each valley into three pieces which together
appear at the six BZ corners.   We choose instead as
a unit cell the parallelogram constructed from the two reciprocal lattice vectors: $\vec a_1 = (2\pi,2\pi/\sqrt{3})$ and
$\vec a_2 = (2\pi,-2\pi/\sqrt{3})$.  The valley points are then $K = (4\pi/3,0)$ and $K' = (8\pi/3,0)$.
Linearizing the tight-binding Hamiltonian around these points leads to:
\begin{eqnarray}
{\cal H}^{Dirac}_{\bk} = v\times\begin{pmatrix} 0 & \pm k_x + i k_y \\ \pm k_x - ik_y & 0\end{pmatrix}
\end{eqnarray}
where the (+)- sign corresponds to the (K')K-valley and $v = \sqrt{3}t/2$ is the quasiparticle velocity.
The difference in sign corresponds to a difference in sign in the pseudospin-chirality\cite{PRL,SSC}
of the Bloch states at the two-valleys which is responsible, as we explain below, for many of the qualitative
features of the LDOS maps.
The unperturbed Matsubara Greens' function is therefore given by
\begin{equation}
G^0(\bk,i\omega) = {1\over \omega^2 + \bk^2}\times \begin{pmatrix} i\omega & \pm k_x + i k_y \\ \pm k_x - ik_y & i\omega\end{pmatrix}.
\end{equation}
For convenience we have rescaled our momentum by the velocity $v$.
We may now evaluate $\Lambda_0$ and $\Lambda_3$ within the linearized model.

\subsection{Intra-valley Scattering}
Here we use the same Greens' function for both the initial and final states ({\em i.e.} the same chirality sign) and arrive at:
\begin{eqnarray}\label{eq:LambdaLin}
\Lambda_{0/3}(\bq) = \;\;\;\;\; \nonumber \\
\int {d^2 k \over (2\pi)^2}{-\omega^2 \pm (k_x+ik_y)(k_x -q_x - ik_y+iq_y) \over (\omega^2 + \bk^2)(\omega^2 + (\bq-\bk)^2)}
\end{eqnarray}
where the (-)+ sign corresponds to ($\Lambda_3$)$\Lambda_0$.
The right hand side of Eq.(\ref{eq:LambdaLin}) can be simplified with the Schwinger/Feynman trick\cite{Peskin}:
\begin{eqnarray}\label{eq:LambdaLin2}
&=&\int_0^1 dx \int {d^2 k \over (2\pi)^2}
{-\omega^2 \pm (k_x+ik_y)(k_x -q_x - ik_y+iq_y)\over (\bk^2 + \omega^2 + (1-x)\bq^2 - 2\bk\cdot\bq(1-x))^2} \nonumber \\
&=&\int_0^1 dx \int {d^2 k \over (2\pi)^2} \; {-\omega^2 \pm (\bk^2 - x(1-x)q^2)\over (\bk^2 + \Delta)^2},
\end{eqnarray}
where $\Delta = \omega^2 + x(1-x)\bq^2$.
Integrating over momenta, we find that the terms with momentum independent numerator yield $(-\omega^2 \mp x(1-x)q^2))/4\pi\Delta$,
whereas the $\bk^2$ numerator terms yields $1+\log(\Delta / \Omega))/4\pi$ where $\Omega$ is an ultraviolet cutoff
that arises in the dimensional regularization scheme\cite{Peskin}.
Integrating over $x$ we find that for $\Lambda_0$ the terms with momentum independent numerators give $-1/4\pi$
so that
\begin{eqnarray}\label{eq:lambda0lin}
\Lambda_0(\bq) &=& -{1\over 4\pi}\left[2-\int dx \log\left({1 \over \omega^2 + x(1-x)\bq^2}\right) \right]\nonumber \\
 &=& {-1 \over 4\pi} \left(2{\cal F}\left({2\omega \over |\bq|}\right) +i\pi + \log\left({\omega^2 \over \Omega^2}\right)\right)
 \end{eqnarray}
 where
 \begin{eqnarray}
 {\cal F}(z) &=& \sqrt{-z^2-1}\arctan{1 \over \sqrt{-z^2-1}}.
\end{eqnarray}
The physical quantity is given by:
\begin{equation}
\delta N_0^A(\bq)= {{\rm sgn}(\omega) \over 2\pi^2} {\rm Im}\left[{\cal F}\left({2i\omega \over |\bq|}\right)\right]
\end{equation}
It is interesting to note that ${\cal F}$ vanishes when $|\bq| = 2\omega$.
This is despite the fact that energy conservation leads to the requirement that the initial and final states be on the same contour of constant energy with radius $\omega$.  This property requires the (-) sign in the $\bk$-independent term to cancel a singularity
at $|\bq| = 2\omega$; the singularity {\em does} appear when the opposite sign is taken in evaluating $\Lambda_3$.  The
absence of this singularity in the intra-band $\Lambda_0$ map is largely responsible for the
qualitative difference between intra-band and inter-band maps.

The physics behind this cancelation can be understood qualitatively
as follows.
We may evaluate the dominant contribution to $\Lambda_{0/3}$ in terms of scattering between eigen states of the unperturbed system.  These states are 2-vectors (pseudospinors) of the form $\Psi_{\bk}^\dagger = (1,\pm e^{-i\phi_{\bk}})$ where
the (-)+ sign corresponds to the (negative) positive energy band and $\phi_{\bk} = \arctan(k_y/k_x)$ near the K valley. Due to energy conservation and to the sharpness of quasiparticles, the dominant scattering events involve on-shell states with energy $\omega$.
The phase space integral for elastic scattering on the energy shell,
\begin{eqnarray}
&&\int d^2 q\delta(\omega-v|\bk|)\delta(\omega-v|k-q|) \nonumber \\
&=&\int dq q d \theta \,  \delta( \omega - [(\omega+vq\cos(\theta))^2+v^2q^2\sin^2(\theta)]^{1/2}) \nonumber \\
&=& \int dq {d(\cos(\theta))\over \sqrt{1-\cos^2(\theta)}}\delta(\cos(\theta)+{q\over 2 \omega}) \nonumber \\
&=& \int dq \; \frac{2\omega \, \Theta(2\omega-vq)}{v \, \sqrt{4\omega^2-v^2q^2}},
\end{eqnarray}
in turns places greatest weight on back-scattering processes with
$\cos(\theta)=-vq/2\omega = -1$. (For each $\bq$ there is one relevant $\bk$ such that $\bk=-(\bk-\bq)$; in the second line we placed the $x$-axis on the $\bk$ direction and replaced $|k|$ by $\omega$ to account for the $\delta$-function.)
However the importance of these processes also
depends on the matrix element of $\sigma_{0/3}$ between the initial and final states:
\begin{eqnarray}
\Lambda_{0/3}(|\bq|=2\omega) \propto \langle \Psi_{\bk} | \sigma_{0/3} |\Psi_{\bk-\bq} \rangle \nonumber \\
=1 \pm e^{i(\phi_{\bk}-\phi_{-\bk})} =1\pm e^{i\pi}
\end{eqnarray}
This factor vanishes when the positive sign is taken for the $\Lambda_0$ case,
because pseudospinors with opposite momentum
in the same valley are orthogonal.  The same effect is responsible
for the Klein-paradox effects in graphene transport properties\cite{Klein} and led to a faster than usual decay of the Friedel oscillations in real space, as calculated by Cheianov and Fal'ko\cite{Cheianov1} and by Bena\cite{Bena}.

As we have just shown, due to pseudospinor-related matrix-element effects the-$\Lambda_{0}$ scattering
amplitude across a Dirac-cone within the same valley vanishes.
In the case of $\Lambda_3$-scattering quite the opposite happens and
the momentum independent terms in the numerator of Eq.(\ref{eq:LambdaLin2}) lead to a divergence along $|\bq| = 2\omega$.
We may therefore neglect the non-singular contribution and write:
\begin{eqnarray}\label{eq:lambda3lin}
\Lambda_3(\bq)&=&\int_0^1 dx \int {d^2 k \over (2\pi)^2}{-\omega^2 + x(1-x)q^2\over (\bk^2 + \Delta)^2}\nonumber \\
&=& {1\over 4\pi}\int_0^1 dx {-\omega^2 + x(1-x)q^2\over \omega^2+x(1-x)q^2}\nonumber \\
&=& {1\over 4\pi}\left[1+2{\cal G}\left({2\omega \over |\bq|} \right)\right]\nonumber \\
 {\cal G}(z) &=& {z^2\over \sqrt{-z^2-1}}\arctan{1 \over \sqrt{-z^2-1}}
\end{eqnarray}
and the physical quantity is:
\begin{eqnarray}
\delta N^A_3(\bq,\omega) = -{{\rm sgn}(\omega)\over 2\pi^2}{\rm Im}\left[{\cal G}\left({2i\omega \over |q|}\right)\right]
\end{eqnarray}
The above results lead to two very different FT LDOS patterns around zero momentum transfer, as shown in Fig.(\ref{fig:zero}).
This may serve as a way to use experimental LDOS maps to distinguish between $\Lambda_0$ and $\Lambda_3$ type potentials in a sample.

\subsection{Inter-valley scattering}
In this subsection we consider the case of momentum transfer $K'-K+\bq$ where $\bq$ is small and therefore the initial and final states are each in the vicinity of a valley, however not the same valley.  This leads to the following expression:
\begin{eqnarray}\label{eq:LambdaLinK}
\Lambda_{0/3}(\bq) =
\int {d^2 k \over (2\pi)^2}{-\omega^2 \pm (k_x+ik_y)(-k_x +q_x - ik_y+iq_y) \over (\omega^2 + \bk^2)(\omega^2 + (\bq-\bk)^2)}
\end{eqnarray}
where the sign in front of $k_x -q_x$ has been reversed from Eq.(\ref{eq:LambdaLin}) since the chirality of the final state at $\bk-\bq$ state is different from the chirality of the initial state at $\bk$.  This sign change leads to an angular dependent intensity peaked along the contour at $|q| = 2\omega$.  Using the Schwinger/Feynman trick with momentum shift as before we arrive at:
\begin{eqnarray}
\Lambda_{0/3}(\bq)&=&\int_0^1 dx \int {d^2 k \over (2\pi)^2}{-\omega^2 \pm x(1-x)(q_x+iq_y)^2\over (\bk^2 + \Delta)^2} \nonumber\\
\Lambda_{0/3}(\bq) &=& {1\over 4\pi}\left[\left(1\pm e^{2i\phi_{\bq}}\right){\cal G}\left({2\omega \over |\bq|}\right)\pm e^{2i\phi_q}\right]
\end{eqnarray}
where the function ${\cal G}(z)$  is defined as before and $\phi_{\bq}=\arctan(q_y/q_x)$.  Note that the physical quantity, $\delta N(\bq,\omega)$, is obtained by analytic continuation of the frequency and has both real and imaginary parts.  On the contour defined by $|\bq|=2\omega$, the real part of $\delta N(\bq)$ has a $\sin^2(\phi_{\bq})$ or $\cos^2(\phi_{\bq})$ angular dependence for $\sigma_0$ and $\sigma_3$ type scattering, respectively. The imaginary part varies as $\pm \sin(2\phi_{\bq})$ along the contour.   These expressions for the two types of scattering differ in their orientation, explaining one of the principle features of the tight-binding model momentum space maps.

\begin{figure}[t]
\includegraphics[width = 0.7\columnwidth]{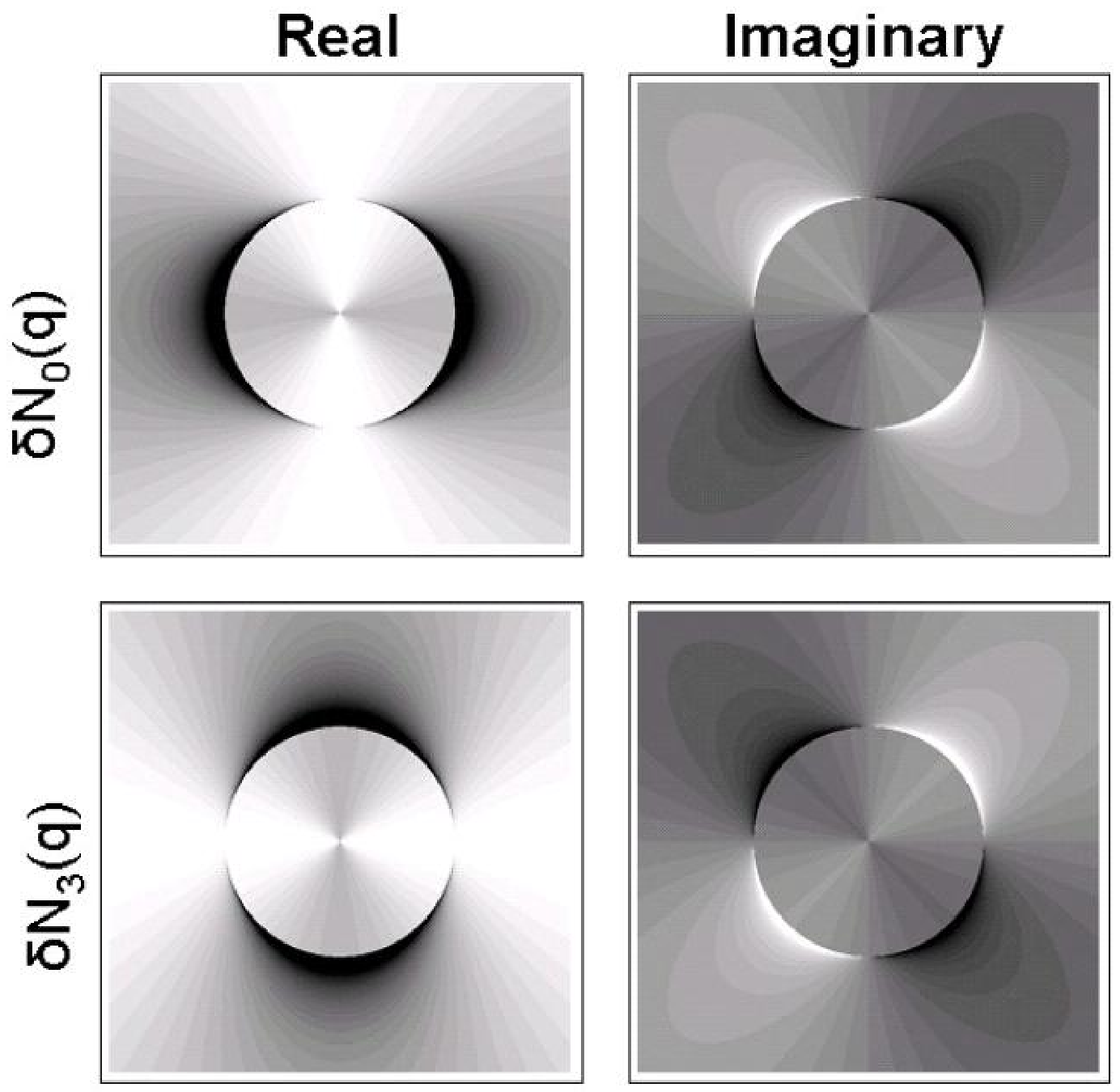}
\caption{The real and imaginary parts of the FT LDOS patterns of $\Lambda_0$ and $\Lambda_3$ type scattering, with momentum transfer around $K'-K = (-4\pi,0)$.Features around other inter-valley scattering vectors are obtained through $2\pi/3$ rotations.}\label{fig:linK}
\end{figure}
\begin{figure}[t]
\includegraphics[width = 0.7\columnwidth]{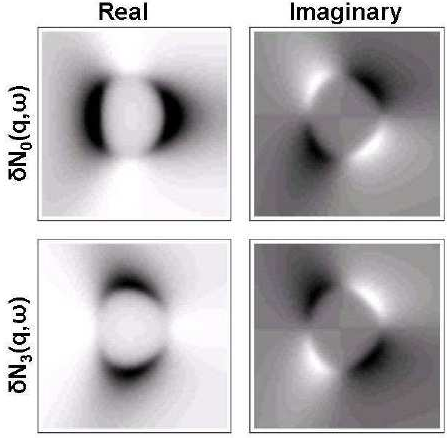}
\caption{FT LDOS - A zoom in around momentum transfer K'-K taken from Fig.(\ref{fig:TB}), to be compared with Fig.(\ref{fig:linK}).}\label{fig:zoom}
\end{figure}
These results are plotted in Fig.(\ref{fig:linK}). Note that the solutions match the ones obtained numerically in the tight binding model (see enlarged features in Fig.(\ref{fig:zoom})).  In order to obtain the full momentum space unit cell, however, one should apply the phase $\exp(i\btau\cdot\bQ)$ to parts of the unit cell which are outside of the Brillouin zone.  This leads to a $2\pi/3$ rotation about $\bq = 0$.

The LDOS modulations at $|\bq| =2\omega$ may be understood through the dominant scattering contribution as before. We may use the same arguments as in the case of intra-valley scattering to show that the dominant process in $\Lambda_{0/3}(K'-K+\bq)$ has $\bk = -(\bk-\bq)+K'-K)$, so that $\bk$ and $\bk-\bq$ are opposite except for the shift in valley.  The eigenvectors are now $\Psi_{\bk}^\dagger = (1, e^{-i\phi_{\bk}})$  and ${\Psi'}_{\bk-\bq}^\dagger = (1, e^{-i\phi'_{\bk-\bq}})$  where $\phi'$ is the angle in the K' valley.  In the K' valley the sign of $k_x$ is reversed (and the chirality of the bands is reversed) so that the angle becomes $\phi'_{\bk} = \pi-\phi_{\bk}$. This leads to \begin{eqnarray}
\Lambda_{0/3}(\bq|=2\omega) \propto \langle \Psi_{\bk} | \sigma_{0/3} |\Psi'_{\bk-\bq} \rangle \nonumber \\
=1 \pm e^{i(\phi_{\bk}-\phi'_{\bk-\bq})} =1\pm e^{2i\phi_{\bq}}.
\end{eqnarray}
The real and imaginary parts of this expression reproduce the angular dependence of Fig.(\ref{fig:linK}).

\subsection{Off-Diagonal disorder}
In the previous sections we have modeled an impurity as an on-site potential.  It is reasonable to assume that impurities may induce a change in the hopping amplitude.  Such an effect can be realized, for example, by stretching of bonds out of the graphene plane.  In this paper we consider a hopping amplitude suppression on three bonds around an impurity on the A site.  This perturbation is non-local, breaks the sublattice symmetry and is described by an off-diagonal potential matrix.

Let us define the following perturbation Hamiltonian:
\begin{eqnarray}
\delta {\cal H}= \sum_{\br}\delta t(\br)\sum_{\bdelta}\left[c^\dagger_{\br}d^{}_{\br+\bdelta}+d^\dagger_{\br+\bdelta}c^{}_{\br}\right]
\end{eqnarray}
where $\delta t(\br)$ is the change in hopping amplitude around the A atom in unit cell $\br$. The operators $c$ and $d$ annihilate an electron on the A and B sublattice respectively.
Writing the perturbation in momentum space and with a pseudospinor vector $\Psi^\dagger = (c^\dagger,d^\dagger)$ we find:
\begin{eqnarray}
\delta {\cal H} = \sum_{\bq}\delta t(\bq)\sum_{\bk} \Psi_{\bk}^\dagger \Gamma(\bk,\bk-\bq)\Psi_{\bk-\bq} \nonumber \\
\Gamma(\bk,\bk-\bq) = \begin{pmatrix} 0 & \gamma_{\bk-\bq} \\ \gamma^*_{\bk} & 0 \end{pmatrix}
\end{eqnarray}
In the Born approximation this leads to $G(\bq,\omega)=\sum_{\bk} G^0(\bk,\omega)\Gamma(\bk,\bk-\bq)G^0(\bk-\bq,\omega)$ such that
\begin{eqnarray}
G_{AA}(\bq,i\omega) = i\omega \sum_{\bk}{|\gamma_{\bk}|^2+|\gamma_{\bk-\bq}|^2 \over (\omega^2+\gamma_{\bk}^2)(\omega^2+\gamma_{\bk-\bq})} \nonumber \\
G_{BB}(\bq,i\omega) = 2i\omega \sum_{\bk}{\gamma_{\bk}^*\gamma_{\bk-\bq}\over(\omega^2+\gamma_{\bk}^2)(\omega^2+\gamma_{\bk-\bq})}
\end{eqnarray}
The numerical and analytic evaluation of the above expressions is similar to that presented before.  We mention, however, that this perturbation is a symmetric function of the bias voltage $\omega$ and therefore may be separated from the on site potential which produces antisymmetric functions.
Since this perturbation is centered around an A atom the LDOS measured on the A sublattices is symmetric in real space and does not have an imaginary part when Fourier transformed.  On the B sublattice, the effects of inter-sublattice scattering is seen and one is able to see both real and imaginary parts with an angular dependent amplitude around inter-valley scattering vectors.  When combining both A and B sublattices the result is very similar to $\Lambda_0(\bq,\omega)$ which was presented in the previous subsections.  In the linear approximation the off diagonal contribution is $\delta N_{off}(\bq,\omega) = 2\omega \delta N_0^A(\bq,\omega)$.

\section{Summary and Discussion}
We have obtained analytic expressions for STM momentum-space LDOS maps in graphene
using the massless Dirac equation model for this material.  We find that
smooth disorder produces features near $\bq=0$ and reciprocal lattice vectors.
The most interesting and surprising feature is the absence of the back-scattering peak
which would be expected on the basis of scattering phase space considerations.
The feature is absent because of the sublattice pseudospin chirality of
Dirac band states which causes disorder induced back-scattering matrix elements
to vanish, the same feature of graphene which helps to enhance its mobility.
Atomic length scale disorder leads to both
$\bq=0$ features and to features near the inter-valley scattering wavevector.
For these features, pseudospin chirality does not cause backscattering matrix elements
to vanish, and instead leads to an interesting angular patterning of the on-shell
peak in the LDOS map.  Our Dirac model analytic results are
in agreement with tight-binding model numerical results.  These results demonstrate the
potential of STM experiments to shed light on the character of disorder in a graphene sample.

In this paper we have not accounted for the influence of electron-electron
interactions on the tunneling DOS of a graphene system.
Indeed the potential of LDOS-map experiments to provide a
high resolution probe of interaction effects in the one-particle
Greens' function of graphene sheets is a major motivation for undertaking these experiments.
On the basis of existing theory\cite{ourarpes,dassarmaarpes}
it appears that electron-electron interactions
alter the quasiparticle velocity, introduce lifetime broadening, and
also under some circumstances, introduce sidebands associated with
plasmon emission.  In the present theory the LDOS-map depends only on
energy relative to the Dirac point, whereas in an interacting system
the map will also depend on the ratio of the energy to the
Fermi energy.  Although the role of interactions must therefore be considered
carefully in analyzing LDOS-map experiments, the disorder and
matrix element considerations explained here will still play a primary role.
One role played by electron-electron interactions will be that of
screening\cite{Nomura,screening}
the external disorder potential, weakening its strength.  The weak-disorder
theory explained here is more likely to be qualitatively correct when the
Fermi level lies away from the Dirac point, so that the Fermi-level
density-of-states is larger, the screened potentials weaker\cite{Nomura,Fuhrer}, and the
Born approximation analysis more accurate.  The Born approximation analysis is
probably invalid at small average carrier densities where the carrier spatial
distribution appears\cite{Yacoby,Adam} highly inhomogeneous.  LDOS-map experiments
should be interesting in both regimes.

\acknowledgements{The authors thank C. Bena for useful comments.  This work was supported in part by the Welch Foundation, by the Army Research Office, by the Natural Sciences and Engineering and Research Council of Canada and by the NRI-SWAN program.}



\begin{thebibliography}{999}
\bibitem{footnote} Note that although the the off-diagonal Greens' functions are gauge dependent, $G^{(1)}_{ii}(\br,\br,\omega)$ is not.
It is important in evaluating $G^{(1)}_{ii}(\br,\br,\omega)$ to be careful that the gauge choices made
in evaluating $G^{(0)}$ are applied
consistently, particular the choice made for the relative phase of Bloch state wavefunction
amplitudes on different sublattices.
\end{thebibliography}
\end{document}